\documentclass[twoside]{LCWS11}
\usepackage[latin1]{inputenc}
\usepackage{graphicx,color}
\usepackage{wrapfig,rotating}
\usepackage{amssymb,amsmath,array}
\usepackage{cite}
\begin{document}
\title{
%%%%   Paper title goes here  %%%%%%%%%%%%%%
A Study of Heavy Higgs Properties\\ at a Multi-TeV $e^+e^-$ Collider} %% 
%***********************************************************************
% AUTHORS INFORMATION AREA
%***********************************************************************
\author{Marco Battaglia$^{1,2,3}$, 
Frederik Bogert$^1$, 
Arnaud Ferrari$^4$, 
Johan Relefors$^4$ 
and Sarah Zalusky$^1$
% DO NOT MODIFY THE FOLLOWING '\vspace' ARGUMENT
\vspace{.3cm}\\
1- University of California at Santa Cruz - Santa Cruz Institute of Particle Physics,  \\
Santa Cruz, CA 95064 - USA\\
2- Lawrence Berkeley National Laboratory, Berkeley, CA 94720 - USA\\
3- CERN, CH-1211 Geneva - Switzerland\\
4- Uppsala University - Department of Physics and Astronomy, 75120 Uppsala - Sweden\\
}
%%***********************************************************************
% END OF AUTHORS INFORMATION AREA
%***********************************************************************

\maketitle

\begin{abstract}
The precise determination of the masses of the neutral and charged heavy Higgs bosons is a 
crucial input for the study of Supersymmetry and its relation with cosmology through dark matter. 
This paper presents a study of $e^+e^- \to H^0A^0$ and $H^+H^-$ production at $\sqrt{s}$=3~TeV.
The analysis is performed with full simulation and reconstruction accounting for beamstrahlung 
effects and the overlay of  $\gamma \gamma \rightarrow {\mathrm{hadrons}}$ events.
Results are presented in terms of the accuracy on the determination of the masses and widths 
of the heavy Higgs bosons in two benchmark scenarios. 
\end{abstract}

\section{Introduction}

The precise determination of the masses and widths of the neutral and charged 
heavy Higgs bosons is an important part of the study of an extended Higgs sector 
in models of new physics. A non-minimal Higgs sector is one of the simplest extension 
of the Standard Model (SM), realised in the Two Higgs Doublet Model, which acquires a special 
relevance and justification in supersymmetric models. In Supersymmetry the detailed study of 
the heavy Higgs sector is crucial to assess the relation between new physics and cosmology,  
through dark matter~\cite{Baltz:2006fm}. A high-energy lepton collider is particularly well 
suited for such a study even in those regions of the parameter space where the sensitivity 
of the LHC becomes marginal. In particular, the pair production processes, 
$e^+e^- \to H^0A^0$ and $e^+e^- \to H^+H^-$ give access to all four heavy Higgs 
states almost up to the kinematical limit~\cite{Coniavitis:2007me,halcc4,halcd}.

\section{Simulation}

In this benchmark study we consider two SUSY models with $M_A$ = 742~GeV and 902~GeV, 
respectively, at $\sqrt{s}$ = 3~TeV and adopt the proposed CLIC beam parameters. 
The model parameters and the masses of the relevant particles are given in Table~\ref{tab:para}. 
The cross sections for $e^+e^- \to H^0A^0$ and $e^+e^- \to H^+H^-$ pair production are given in 
Table~\ref{tab:xsec}. The properties of the heavy Higgs sector are largely independent on the details 
of supersymmetric models and chosen mostly depend on the values of $M_A$ and $\tan \beta$. 
It is interesting to observe that the chosen values of these parameters are compatible with the 
current LHC data and with a $\sim$125~GeV lightest supersymmetric Higgs boson, $h^0$, in the 
general MSSM. In particular, the scenario~1 is compatible with a non-observation of SUSY signals in 
the channels with missing transverse energy by the end of the 8~TeV LHC run and 123$<M_h<$127~GeV 
in the MSSM~\cite{Arbey:2011aa}.
\begin{table}
\begin{center}
\caption{Model parameters and Higgs boson masses of the two benchmark points studied.}
\begin{tabular}{|l|c|c|}
\hline
                 & Scenario 1 & Scenario 2 \\ \hline
Mass Parameters (GeV)  & $M_1$ = 780~$M_2$ = 940~$M_3$ = 540 & $m_0$ = 800~$m_{1/2}$ = 966 \\
$\tan \beta$, $A_0$, sign($\mu$) & 24,~-750,~+ & 51,~0,~- \\
$M_{A}$, $M_{H^0}$, $M_{H^{\pm}}$ (GeV) & 902.6,~902.4,~906.3 & 742.8,~742.0,~747.6 \\
\hline
\end{tabular}
\label{tab:para}
\end{center}
\end{table}
Signal events are generated with {\tt Isasugra~7.69}~\cite{Paige:2003mg} and 
{\tt Pythia~6.215}~\cite{Sjostrand:2006za}.
Taking into account the CLIC luminosity spectrum, the predicted signal cross sections for $H^0A^0$ and $H^+H^-$ are 
respectively 0.7 (0.4)~fb and 1.6 (1.1)~fb in the scenario 1 (2). 
The dominant decay modes are $H^0 \to b \bar b$, $A^0 \to b \bar b$ and $H^+ \to t \bar b$ leading to the 
$b \bar b b \bar b$ and $t \bar b \bar t b$ final states.
\begin{table}[h!]
\begin{center}
\caption{Summary of the processes considered in this study with their production cross section and the event generator used.}
\begin{tabular}{|l|c|c|}
\hline
Process     & $\sigma$ (fb) & Generator \\
\hline
$H^0A^0$    &  ~~0.7 / ~~0.4   & ISASUGRA~7.69+PYTHIA~6.215   \\
$H^+H^-$    &  ~~1.6 / ~~1.1    & ISASUGRA~7.69+PYTHIA~6.215   \\
\hline
Inclusive   & ~84.9 / ~77.1   & ISASUGRA~7.69+ \\
SUSY        &          & PYTHIA~6.215   \\
$W^+W^-$    & 728.2    & PYTHIA~6.215   \\
$Z^0Z^0$    & ~54.8    & PYTHIA~6.215   \\
$t \bar t$  & ~30.2    & PYTHIA~6.215   \\
$b \bar b b \bar b$& ~~5.8 & WHIZARD   \\
$W^+W^-Z^0$ & ~32.8    & CompHEP+PYTHIA~6.215   \\
$Z^0Z^0Z^0$ & ~~0.5    & CompHEP+PYTHIA~6.215   \\
\hline
\end{tabular}
\label{tab:xsec}
\end{center}
\end{table}   
The main SM background processes are generated with {\tt Pythia~6.215} and {\tt WHIZARD}. 
In addition, the irreducible inclusive 
$b \bar b b \bar b$ SM background is generated with {\tt CompHep}~\cite{Comphep} at the 
parton level and subsequently hadronised with {\tt Pythia}.
We assume the beams to be unpolarised. Events are processed through full detector simulation using the 
{\tt Geant-4}-based {\tt Mokka}~\cite{MoradeFreitas:2004sq} program and reconstructed with
{\tt Marlin}-based~\cite{Gaede:2006pj} processors assuming a version of the ILD 
detector~\cite{ild}, modified for physics at CLIC.
Particle tracks are reconstructed by the combination of a Time Projection 
Chamber and a pixellated Si Vertex Tracker. The momentum resolution is 
$\delta p/p^2$ = 2~$\times$10$^{-5}$~GeV$^{-1}$ and the impact parameter 
resolution is $\sigma_{R-\Phi}$ = ($2.5~\oplus~\frac{21}{p_t {\mathrm{(GeV)}}})$~$\mu$m.
The parton energy is reconstructed using the {\tt Pandora} particle 
flow algorithm~\cite{Thomson:2009rp}. Performances for our signal events are discussed in the next session.
For the $\gamma \gamma \rightarrow {\mathrm{hadrons}}$ background simulation 
two-photon events are generated with the {\tt Guinea~Pig} program~\cite{c:thesis}
using the nominal CLIC CDR beam parameters at $\sqrt{s}$ = 3~TeV where the hadronic 
background cross section is modelled following~\cite{c:had0}. 
The energies of two colliding photons are stored with their corresponding probability and passed to 
{\tt Pythia} for the generation of the hadronic events. On average, there are 
3.3 $\gamma \gamma \rightarrow {\mathrm{hadrons}}$ events per bunch crossing (BX) with $M_{\gamma \gamma} >$ 3~GeV.  
These are passed through the same {\tt Mokka} full detector simulation and overlayed to the particles 
originating from the primary $e^+e^-$ interaction at the reconstruction stage, using the event overlay 
feature in the {\tt lcio} persistency package~\cite{Gaede:2005zz}. For this analysis, 
the $\gamma \gamma$ background is overlayed only to the signal $H^0A^0$ and $H^+H^-$ events, in order 
to study its effect on the signal reconstruction and to the peaking background from the cross-feed of $H^+H^-$ 
events in the $H^0A^0$ analysis and vice versa.

\section{Event Reconstruction and Signal Selection}

The event reconstruction and selection is based on the identification of four 
heavy parton final states in spherical events with large visible energy and equal 
di-parton invariant masses. Details of the $bbbb$ analysis can be found in~\cite{halcd}. 
Most of the analysis criteria are common to both the 
$bbbb$ and the $tbtb$ channels. The analysis starts with a cut-based event pre-selection. 
Jet clustering is applied to pre-selected events, followed by $b$ and $t$ identification. 
A kinematic fitting is performed to improve the di-jet invariant mass resolution, mitigate 
the impact of machine-induced backgrounds on the parton energy resolution and reject
remaining physics backgrounds. 
  
In order to reject particles which are poorly reconstructed or likely to originate 
from underlying $\gamma \gamma$ events, a set of minimal quality cuts are applied. Only 
particles with transverse momentum $p_t>$0.95~GeV are considered, charged particles are 
also required to have at least 12 hits in the tracking detectors and relative momentum 
uncertainty $\delta p/p<$1. 
The event selection proceeds as follows. First multi-jet hadronic events with large 
visible energy are selected. We require events to have at least 50 charged 
particles, total reconstructed energy exceeding 2.3~TeV, event thrust between 0.62 and 0.91, 
sphericity larger than 0.04 and smaller than 0.75, transverse energy exceeding 1.3~TeV and 
3~$\le N_{\mathrm{jets}} \le$~5, where $N_{\mathrm{jets}}$ is the natural number of jets 
reconstructed using the Durham clustering algorithm~\cite{Catani:1991hj} with $y_{cut}$=0.0025. 
These cuts remove all the SUSY events with missing energy and the 
$e^+e^- \to f \bar f$ events. For events fulfilling these criteria, we perform the final 
jet reconstruction using the anti-kt clustering algorithm in cylindrical 
coordinates~\cite{Cacciari:2008gp}, implemented in the {\tt FastJet} package, ported as a custom 
processor in the {\tt Marlin} analysis framework. The choice of cylindrical coordinates is 
optimal since the $\gamma \gamma \to {\mathrm{hadrons}}$ events are forward boosted, similarly 
to the underlying events in $pp$ collisions at the LHC, for which the anti-kt clustering has been 
conceived and optimised.
For each event, the minimum $R$ value at which the event has exactly four jets with energies in 
excess of 150~GeV is used for the clustering.  The di-jet invariant mass is computed from pairing 
these jets. Since there are three possible permutations for pairing the four energetic jets and the 
pair-produced bosons are expected to be (almost) degenerate in mass, we take the combination 
minimising the difference $\Delta M$ of the two di-jet invariant masses and require 
$|\Delta M| <$~160~GeV, 150~GeV for the $H^0A^0$ and $H^+H^-$, respectively. Since the signal events 
are predominantly produced in the central region while the $\gamma \gamma$ events and most of the 
SM background processes are forward peaked, we only accept events for which the jet with 
the minimum polar angle, $\theta$, has  $|\cos \theta| <$ 0.92.
The single most effective event selection to separate signal events from the SM backgrounds is 
$b$-tagging, since the signal contains four $b$ hadrons. The irreducible SM $b \bar b b \bar b$ background 
has a cross section of only 0.5~fb and is effectively reduced by the equal di-jet mass constrain and 
kinematic fitting. The $b$-tagging procedure is based on the response of the vertexing variables of the 
{\tt ZVTOP} program~\cite{Bailey:2009ui}. This uses the kinematics and topology of the secondary 
particles in the jet. For this analysis, the vertex variables are supplemented by the corresponding kinematic 
observables for the secondary system reconstructed based on the charged particle impact parameters instead 
of the topological vertexing, when {\tt ZVTOP} does not return any secondary vertex. This procedure 
increases the efficiency for $b$ jets at the higher end of the kinematic spectrum in signal events. 
Tagging observables are combined into a single discriminating variable using the boosted decision tree 
method with the {\tt TMVA} package~\cite{Hocker:2007ht}. 
In the case of charged Higgs bosons, top tagging is also performed. First the event is reconstructed as a four 
jet event and jets are tested for their compatibility with the top mass. Then a de-clustering procedure is 
applied to the jets to study possible jet substructure arising from the $t \to W b \to q \bar q' b$ decay. 
This follows the procedure originally developed for identifying highly boosted top quarks at the 
LHC~\cite{lhc-top,cms-top}. 

In order to improve the di-jet mass resolution, we apply a constrained kinematic fit.
We use the port of the {\tt Pufitc} kinematic fit algorithm~\cite{pufitc} to the 
{\sc Marlin} framework. {\tt Pufitc} was originally developed for $W^+W^-$ reconstruction 
in DELPHI at LEP-2 and it has been successfully applied for the reconstruction of simulated 
linear collider events at lower energies~\cite{halcc4}. The kinematic fit adjusts the momenta 
of the four jets as $p_F = a p_M + b p_B + c p_C$, where $p_M$ is the jet momentum from 
particle flow, $p_B$ and $p_C$ are unit vectors transverse to $p_M$ and to each other, 
$a$, $b$ and $c$ free parameters. For these analyses, we impose the constraints $p_x$ = $p_y$ 
= 0, $E \pm |p_z| = \sqrt{s}$ and $M_{jj1} = M_{jj2}$, where the second condition accounts for 
beamstrahlung photons radiated along the beam axis. We only accept events with a 
kinematic fit $\chi^2 <$~5. After the kinematic fit, the relative jet energy resolution 
$\mathrm{RMS_{90}}/E_{\mathrm{jet}}$ for $b$ jets is 0.075$\pm$0.002 and 
0.096$\pm$0.002 without and with $\gamma \gamma$ background overlayed, respectively.
The use of a kinematic fit also mitigates the effect of the overlayed $\gamma \gamma$ 
events on the di-jet mass resolution. Since we do impose the nominal centre-of-mass energy, 
allowing for beamstrahlung, jet energies are rescaled in the fit to be consistent with that 
of $\sqrt{s}$. The di-jet invariant mass resolution changes from (38$\pm$4)~GeV, for the raw 
particle flow objects, to (29$\pm$3)~GeV, using the kinematic fit, and to (18$\pm$2)~GeV, imposing 
the equal mass constrain, in absence of $\gamma \gamma$ background. Using the kinematic fit and 
equal mass constrain we obtain (19.3$\pm$3.0)~GeV with the $\gamma \gamma$ background overlayed. 
The use of the equal mass constrain is justified in the case of the $e^+e^- \to H^0A^0$ channel 
in SUSY models. In fact, the mass splitting between the $A^0$ and $H^0$ bosons is always smaller than 
both the natural widths and the experimental di-jet mass resolution, in particular in the region of 
MSSM parameters where their determination is most crucial to fix the relic dark matter density 
$\Omega_{\chi} h^2$.
This has been verified through a dedicated study using {\tt FeynHiggs}~\cite{Hahn:2009zz}. The few cases 
in which the mass splitting becomes large are also characterised by very large natural widths of the 
two bosons.

\section{Results}

\begin{figure}[h!]
\begin{center}
\begin{tabular}{cc}
\includegraphics[width=0.35\textwidth]{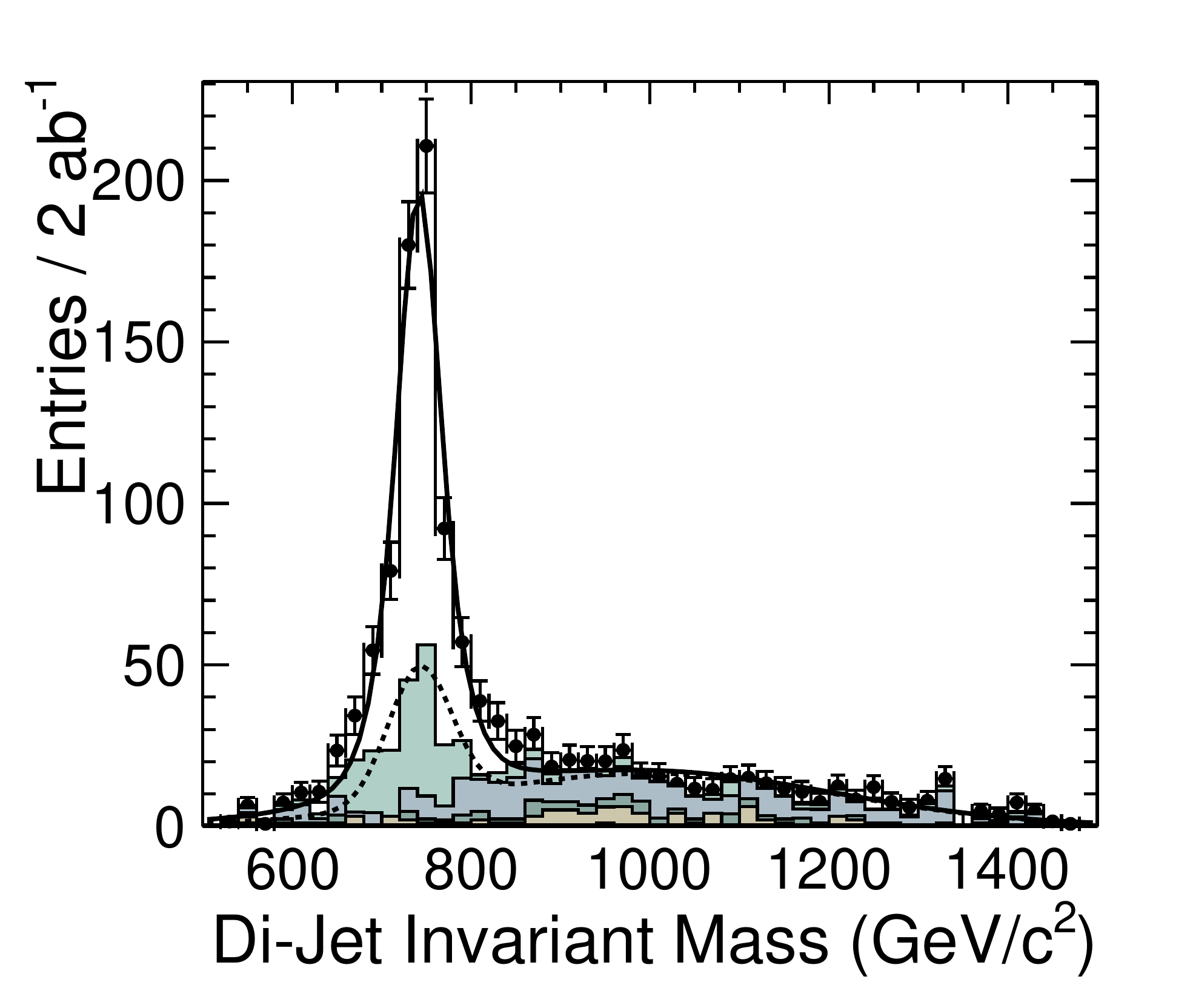} &
\includegraphics[width=0.35\textwidth]{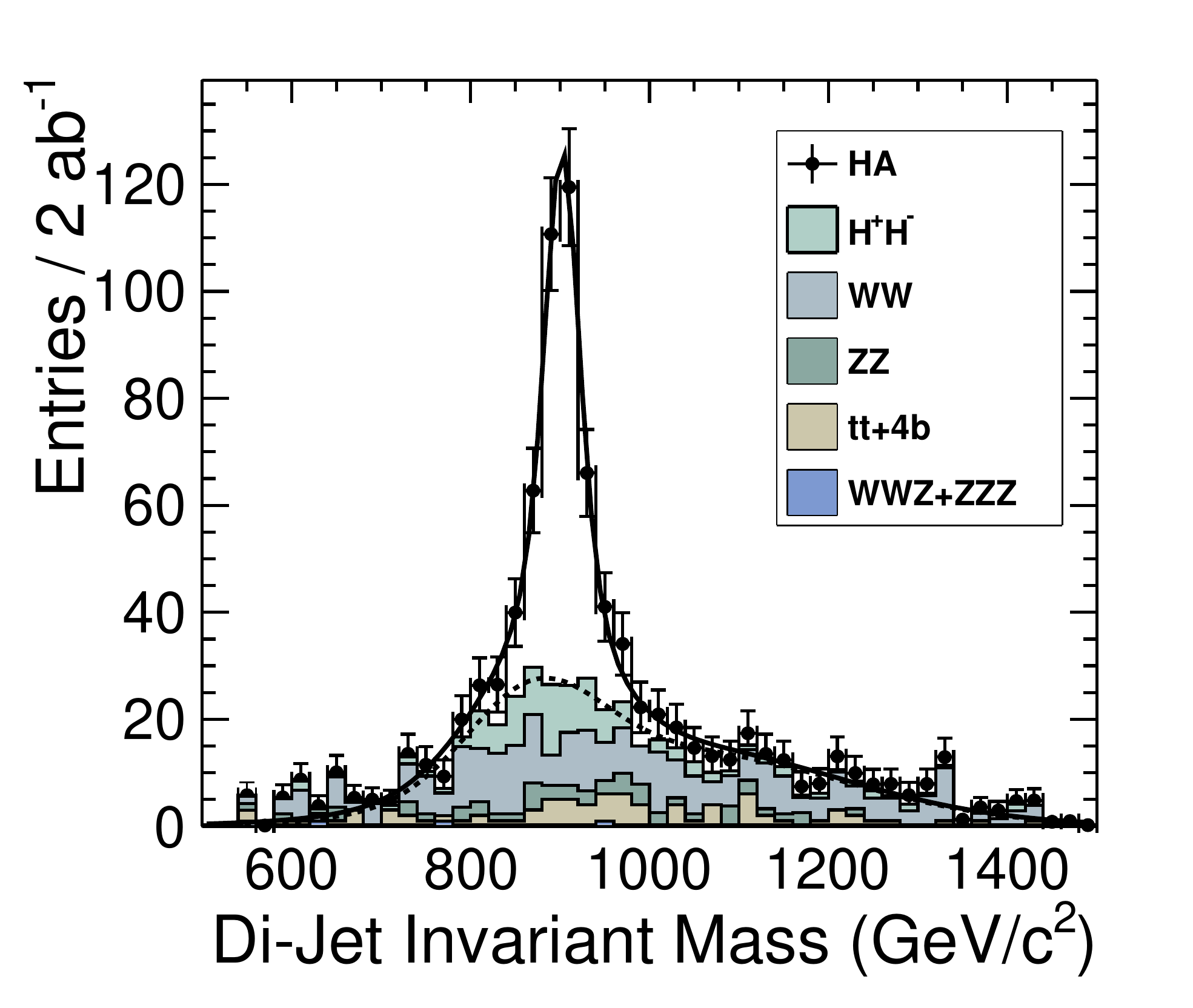} \\
\includegraphics[width=0.35\textwidth]{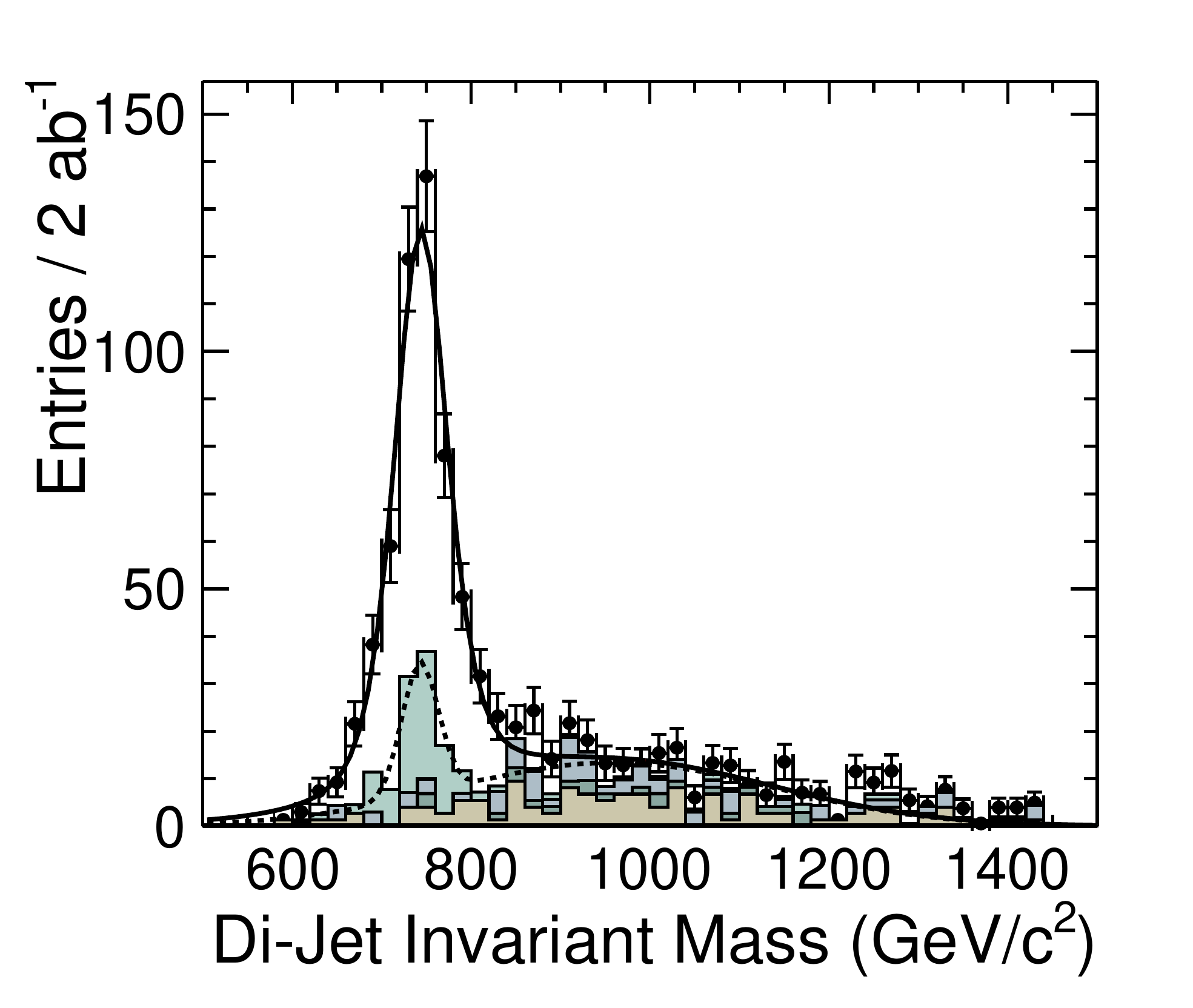} &
\includegraphics[width=0.35\textwidth]{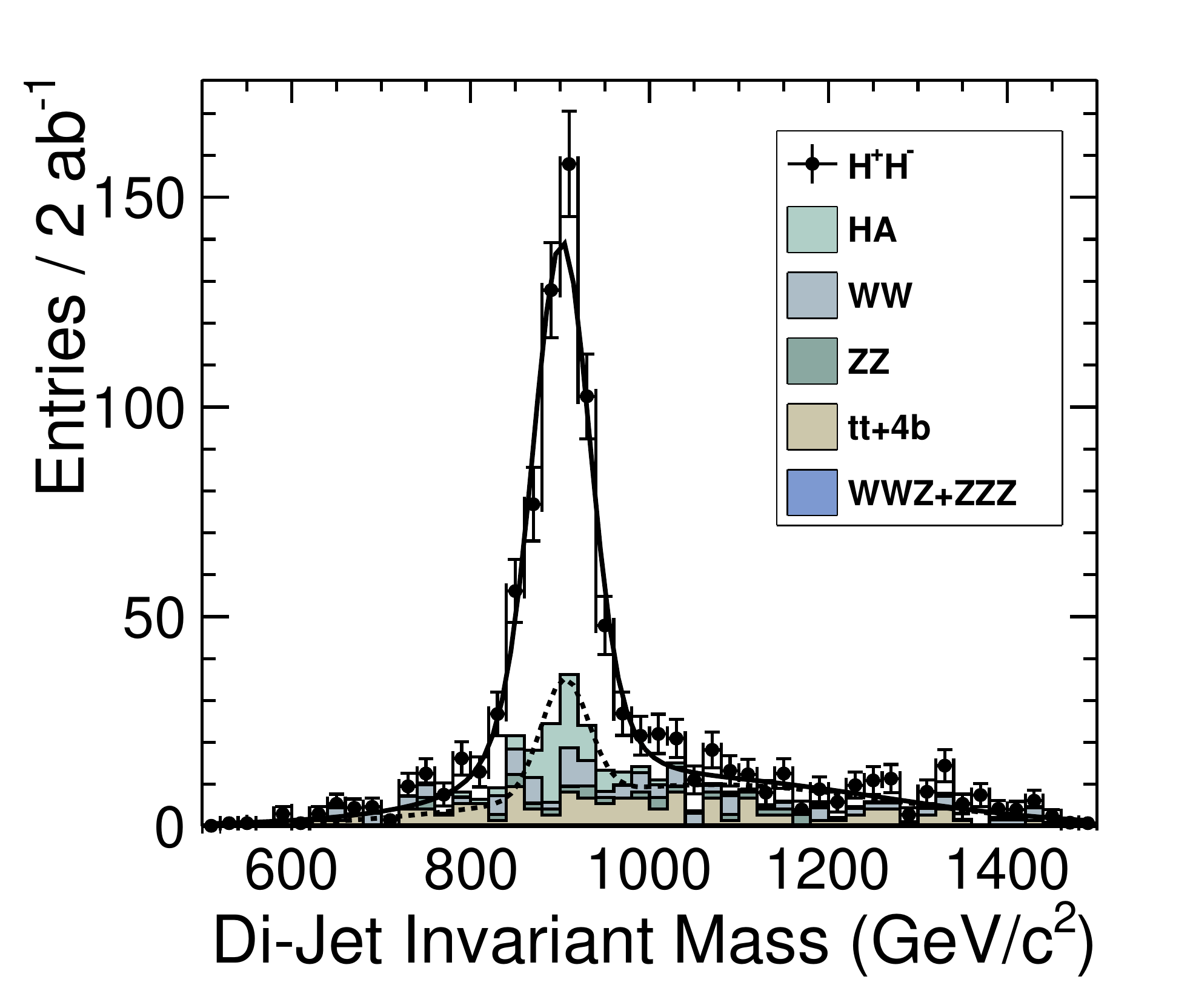} \\
\end{tabular}
\vspace*{-0.5cm}
\caption{Di-jet invariant mass distribution of fully simulated and reconstructed events in the 
channel $e^+e^- \to b \bar b b \bar b$ (upper row) and  $e^+e^- \to t \bar b \bar t b$ (lower row).
The contributions from the different processes are shown. The dotted line shows the fitted background 
distribution which includes the peaking contribution of the cross feeds from the $H^0A^0$ and $H^+H^-$ 
processes. The continuous line shows the global fit including the signal.}
\label{fig:massfit}
\end{center}
\end{figure}

The di-jet invariant mass distributions for the $b \bar b b \bar b$ and $t \bar b \bar t b$ final states 
are shown in Figure~\ref{fig:massfit}. 
\begin{table}[h!]
\begin{center}
\caption{Summary of the fit results for the masses and widths for benchmark scenario~1, 
without and with $\gamma \gamma$ background overlay. The quoted uncertainties are statistical only.}
\begin{tabular}{|c|c|c|c|}
\hline
          & State        & Mass  & Width    \\
          &              & (GeV) & (GeV)    \\
\hline
No $\gamma \gamma$  & $A^0$/$H^0$  & 742.7$\pm$1.4 & 21.7$\pm$3.3  \\
No $\gamma \gamma$  & $H^{\pm}$    & 744.3$\pm$2.0 & 17.0$\pm$4.7  \\
$\gamma \gamma$ (20 BX)   & $A^0$/$H^0$  & 743.7$\pm$1.7 & 22.2$\pm$3.8  \\
$\gamma \gamma$ (20 BX) & $H^{\pm}$    & 746.9$\pm$2.1 & 21.4$\pm$4.9  \\
\hline
\end{tabular}
\label{tab:h1}
\end{center}
\end{table}
\begin{table}[h!]
\begin{center}
\caption{Summary of the fit results for the masses and widths for benchmark scenario 2,  
without and with $\gamma \gamma$ background overlay. The quoted uncertainties are statistical only.}
\begin{tabular}{|c|c|c|c|}
\hline
          & State        & Mass  & Width    \\
          &              & (GeV) & (GeV)    \\
\hline
No $\gamma \gamma$  & $A^0$/$H^0$  & 902.1$\pm$1.9 & 21.4$\pm$5.0  \\
No $\gamma \gamma$  & $H^{\pm}$    & 901.4$\pm$1.9 & 18.9$\pm$4.4  \\
$\gamma \gamma$ (20 BX) & $A^0$/$H^0$  & 904.5$\pm$2.8 & 20.6$\pm$6.3  \\
$\gamma \gamma$ (20 BX) & $H^{\pm}$    & 902.6$\pm$2.4 & 20.2$\pm$5.4  \\
\hline
\end{tabular}
\label{tab:h2}
\end{center}
\end{table}
The masses and widths of the 
heavy bosons are extracted from a multi-parameter fit to these distributions. In order to assess the 
effect of the $\gamma \gamma \to {\mathrm{hadrons}}$ background on the accuracy of the mass and width 
reconstruction, we repeat the analysis assuming first no $\gamma \gamma$ background and then we overlay 
the background assuming an effective 10~ns time stamping capability of the detector, corresponding to the 
integration of 20 bunch crossings (BX). Results are summarised in Tables~\ref{tab:h1} and \ref{tab:h2}.

We observe that the heavy Higgs mass can be measured to a relative statistical accuracy of better than 
0.5\%. This accuracy is preserved in the presence of $\gamma \gamma$ background by applying the anti-kt jet 
clustering and kinematic fitting, provided the global detector stamping resolution is $\le$10~ns. 
The width can be measured to a relative accuracy of 0.20-0.30. These measurements are important 
both for constraining the model parameters through $M_A$ and $\tan \beta$ and for assessing the 
contribution of the $\tilde \chi^0 \tilde \chi^0 \to A^0 \to b \bar b$ pole annihilation process 
in setting the relic dark matter density in the Universe~\cite{Baltz:2006fm}.
These accuracies ensures that the heavy Higgs mass contribution to the statistical precision in the 
extraction of $\Omega_{\chi} h^2$ in a generic MSSM model, with dark matter annihilation through the 
$A^0$ pole, is of order of 0.10. In the specific case of scenario~2, where $\tan \beta$ is large 
and $M_{A}$=742~GeV, these constraints on $\Gamma_{A^0}$ and $\Gamma_{H^{\pm}}$ also yield a relative 
statistical accuracy on the determination of $\tan \beta$ to $\sim$0.20.

\section{Conclusions}

A study of $H^0A^0$ and $H^+H^-$ production in $\sqrt{s}$ = 3~TeV $e^+e^-$ collisions performed on fully 
simulated and reconstructed events with realistic backgrounds shows that the mass and widths of these bosons 
can be determined with a relative accuracy of 0.05 and 0.2-0.30, respectively, provided time stamping with 
a resolution $\le$10~ns can be achieved in the detector.

\begin{footnotesize}

\end{footnotesize}

\end{document}